\documentclass[aps,prd,preprint,floatfix,11pt,showpacs,showkeys,tightenlines,nofootinbib,longbibliography,a4paper]{revtex4-1}
\usepackage[utf8]{inputenc}
\usepackage{soul}
\usepackage[table]{xcolor}
\usepackage[compat=1.0.0]{tikz-feynman}
\usepackage{amsfonts}
\usepackage{amsmath}
\usepackage{array} 
\usepackage[markup=nocolor]{changes}
\usepackage{graphicx} 
\usepackage{hyperref}
\usepackage{multirow}
\usepackage{diagbox} 
\usepackage{commath}
\usepackage{url}
\usepackage{xspace}
\usepackage{setspace}
\usepackage{braket}
\usepackage{graphics}
\usepackage[parfill]{parskip}
\usepackage{amssymb}
\usepackage{pgfplots}
\hypersetup{pdfstartview=FitV,colorlinks=true,linkcolor=blue,citecolor=blue, filecolor=black,urlcolor=blue}

\newcommand{\tr}{\mathrm{Tr}}
\newcommand{\be}{\begin{equation}}
\newcommand{\ee}{\end{equation}}
\newcommand{\bea}{\begin{eqnarray}}
\newcommand{\eea}{\end{eqnarray}}

\newcommand{\msbar}{\overline{\text{MS}}}
\definecolor{lime}{HTML}{A6CE39}
\DeclareRobustCommand{\orcidicon}{\hspace{-2mm}
	\begin{tikzpicture}
	\draw[lime, fill=lime] (0,0) 
	circle [radius=0.16] 
	node[white] at (-0.007,-0.007) {{\fontfamily{qag}\selectfont \tiny \,ID}};
	\draw[white, fill=white] (-0.067,0.095) 
	circle [radius=0.005];
	\end{tikzpicture}
	\hspace{-3.5mm}
}

\foreach \x in {A, ..., Z}{\expandafter\xdef\csname orcid\x\endcsname{\noexpand\href{https://orcid.org/\csname orcidauthor\x\endcsname}
			{\noexpand\orcidicon}}
}
\usepackage{calc}
\newlength{\depthofsumsign}
\makeatletter
\newcommand*{\DivideLengths}[2]{%
  \strip@pt\dimexpr\number\numexpr\number\dimexpr#1\relax*65536/\number\dimexpr#2\relax\relax sp\relax
}
\makeatother

\begin{document}
\title{One-loop renormalization and  $\boldsymbol{\rho}$ parameter in the Georgi-Machacek model}
\author{Debtosh Chowdhury\orcidA{}$^{,1,}${\footnote{
Electronic address: debtoshc@iitk.ac.in}},
Anirban Kundu\orcidD{}$^{,2,}${\footnote{
Electronic address: akphy@caluniv.ac.in}},
Poulami Mondal\orcidB{}$^{,1,3,}${\footnote{
Electronic address: poulami.mondal@tifr.res.in}}, \\
Subrata Samanta\orcidC{}$^{,1,}${\footnote{
Electronic address: samantaphy20@iitk.ac.in}}, 
Aaryan Srivastava\orcidE{}$^{,1,}${\footnote{
Electronic address: aaryans20@iitk.ac.in}}
 \\[8pt]
$^1$\textit{Department of Physics, Indian Institute\\ of Technology Kanpur, Kanpur 208016, India}\\[5pt]
$^2$\textit{Department of Physics, University of Calcutta, \\
92 Acharya Prafulla Chandra Road, Kolkata 700009, India}\\[5pt]
$^3$\textit{Department of Theoretical Physics, Tata Institute of Fundamental Research, Mumbai 400005, India}
}
\preprint{TIFR/TH/26-2}
\begin{abstract}
We study the one-loop renormalization of the Georgi-Machacek model. At one loop, the renormalization of the model is phenomenologically important when triggered by operators that are absent at the tree level due to the global $SU(2)_R$ symmetry. By computing all the tree-level parameters from the standard input parameters $\alpha_e$, $G_\mu$, and $m_Z$, we show the ultraviolet divergent nature of the electroweak $\rho$ parameter when one-loop corrections are incorporated. In this model, four input parameters are required to completely parametrize the electroweak precision observables at one loop. We study the quantitative impact of the model parameters on the one-loop corrections to the $\rho$ parameter. At one loop, the $\rho$ parameter shows a mild dependence on the mass differences between the custodial fiveplet and the heavy custodial singlet, and mainly depends on the ratio of the doublet and triplet vacuum expectation values, and on the mixing angle  between the custodial singlet CP-even scalars.
\end{abstract}
\maketitle
\cleardoublepage
\section{Introduction}\label{sec:intro}
The electroweak (EW) observables are measured with high precision in both low-energy and high-energy experiments~\cite{ALEPH:2013dgf,LHCb:2021bjt,ATLAS:2024erm,ParticleDataGroup:2024cfk,CMS:2024lrd}. One of the key EW precision parameters inferred from these observables is the $\rho$ parameter, which quantifies the custodial $SU(2)_V$ breaking in the theory involving scalar fields. If the custodial $SU(2)_V$ symmetry is exact,  $\rho=1$. The global EW fit value of the $\rho$ parameter is very close to unity~\cite{ParticleDataGroup:2024cfk,Haller:2018nnx,deBlas:2021wap,deBlas:2022hdk}. Thus, the precision experimental data indirectly indicate that the electroweak symmetry breaking (EWSB) dynamics must possess an
approximate custodial $SU(2)_V$ symmetry in the gauge boson sector, ensuring small corrections to the relation $\rho=1$. In the Standard Model (SM) with a single Higgs doublet, the relation $m_W^2=c_\theta^2m_Z^2$ (translates into $\rho=1$) holds only at tree-level, where $c_\theta$ denotes the cosine of the weak mixing angle ($\theta_W$) and $m_{W},m_Z$ are the
electroweak gauge boson masses. Electroweak quantum corrections, arising from the top–bottom doublet and the Higgs boson, modify this relation~\cite{Veltman:1977kh,PhysRevD.22.971,Hollik:1988ii,Degrassi:1990tu,Denner:1991kt}. Since these electroweak quantum corrections in the SM are precisely calculable~\cite{Bardin:1986fi,Awramik:2003rn,Dubovyk:2019szj,Chen:2020xzx,Freitas:2020kcn}, any additional deviation of the $\rho$ parameter value from the SM prediction can be used to constrain possible new physics effects. Beyond the SM theories with additional scalar multiplets will contribute to the $\rho$ parameter~\cite{Kennedy:1988rt,Kennedy:1988sn}. These contributions can appear at the tree-level itself, if the scalar multiplets develop a vacuum expectation value (VEV) after EWSB~\cite{Passarino:1990xx, Lynn:1990zk}. For example, the SM with an additional scalar triplet under $SU(2)_L$ has a tree-level contribution to the $\rho$ parameter, leading to $\rho\neq1$ at the tree level~\cite{Passarino:1989py,Passarino:1990nu,Lynn:1990zk,Arkani-Hamed:2002ikv, Kanemura:2012rs,Drauksas:2023isd}. To remain consistent with the EW fit results, this tree-level contribution must be of the order of standard quantum corrections. As a consequence, the triplet VEV must be very small. In contrast, the Georgi–Machacek (GM) model~\cite{Georgi:1985nv,Chanowitz:1985ug}, which includes both a complex and a real scalar triplet under 
$SU(2)_L$ in the SM, preserves $\rho=1$ at tree level due to an approximate global $SU(2)_R$ symmetry in the Higgs Lagrangian, analogous to the SM. Consequently, this model allows for sizable triplet VEVs while remaining consistent with electroweak precision data. Gunion, Vega, and Wudka analyzed the basic phenomenology of this model at tree level~\cite{Gunion:1989ci}, and subsequently extended it to the one-loop level in~\cite{Gunion:1990dt}. However, even if the Higgs Lagrangian is $SU(2)_R$ symmetric at tree level, hypercharge gauge interactions reintroduce $SU(2)_R$-breaking terms at the loop level because of the presence of two triplets with different hypercharge quantum numbers. The one-loop phenomenology of the GM model is particularly intriguing, as the $SU(2)_R$ symmetry in the Higgs Lagrangian must be broken at one-loop in order to renormalize the model~\cite{Gunion:1990dt,Englert:2013zpa,Hartling:2014aga,Blasi:2017xmc,Chiang:2018xpl,Keeshan:2018ypw,Mondal:2022xdy,Chowdhury:2024mfu,Mondal:2025tzi}.

The GM model features an extended scalar sector comprising two singly charged, one doubly charged, and four neutral physical scalars. These scalars form a fiveplet, a triplet, and two singlets by their transformation properties under the custodial $SU(2)_V$ symmetry. Each multiplet has a definite custodial $SU(2)_V$ quantum number. As a result, the members of different multiplets do not mix with each other at the tree level. However, such mixings are generated at the one-loop level through the hypercharge gauge interactions, leading to ultraviolet divergent contributions.\footnote{Note that, the custodial $SU(2)_V$ symmetry is an exact symmetry in the absence of hypercharge gauge interactions.} To cancel these divergences in the renormalization procedure, $SU(2)_R$-breaking counter-terms must be introduced. In Refs.~\cite{Gunion:1990dt,Hartling:2014aga}, it is mentioned that the GM model is renormalizable by adding $SU(2)_R$-breaking counter-terms in the scalar potential, and the finite part can be adjusted to compensate the one-loop contribution to the $\rho$ parameter. However, within the modified minimal subtraction ($\overline{\text{MS}}$) scheme, when renormalization is carried out analogously to the SM using the electromagnetic fine structure constant ($\alpha_e$),  the Fermi constant ($G_\mu$), and the $Z$-boson mass ($m_Z$) as input parameters, we find that the one-loop correction to the $\rho$ parameter remains ultraviolet divergent, even after including all $SU(2)_R$ breaking counter-terms in the scalar potential. This is due to the fact that in the presence of two triplets with unequal hypercharge, the hypercharge gauge interactions at the one-loop contribute differently to the $W$ and $Z$ boson self-energies, leading to an ultraviolet divergent contribution to $\rho$. The full one-loop calculations, in the GM model, require an extension of the renormalization procedure. The counter-term for the quantity $s_\theta^2\;(=1-c_\theta^2)$ is an independent parameter. Its value can be chosen from low-energy experiments to compensate precisely for the loop (quantum) corrections to neutrino-electron scattering amplitudes. This is similar to the renormalization procedure for the model with one triplet scalar, where the effective lepton-mixing angle at the $Z$-peak is taken as an input parameter~\cite{Blank:1997qa,Chen:2003fm,Kanemura:2012rs}.

In this paper, we discuss the renormalization procedure for the Higgs sector and the electroweak $\rho$ parameter at the one-loop level in the GM model. Following~\cite{ParticleDataGroup:2016lqr,ParticleDataGroup:2024cfk}, we choose the parameter $\rho_0$, defined as $\rho_0\equiv m_W^2/(m_Z^2c_\theta^2\rho)$, to describe the new-physics effects. Here, $\rho$ is computed assuming the validity of the SM. Hence, the relation $\rho_0=1$ remains valid to all orders in the perturbative expansion in the SM. In the GM model, $\rho_0=1$ at the tree level, hence the gauge boson-fermion sector is parameterized by three free parameters at the tree level. However, at the one-loop level, $\rho_0\neq1$, and is ultraviolet divergent. As a result, four independent input parameters, not three, are required to completely parametrize the electroweak precision observables in the GM model. We can choose $\rho_0$ (or $s_\theta^2$, or $m_W$) itself as the additional input parameter, together with the standard inputs $\alpha_e$, $G_\mu$, and $m_Z$, following the standard $\msbar$ scheme~\cite{Passarino:1990xx}. Our results show that, at one-loop, the $\rho_0$ parameter depends on the triplet VEV, on the mixing angle between the custodial singlet charge-parity (CP)-even scalars, and logarithmically on the masses of the beyond the SM (BSM) scalar states in the GM model.

In the following, we first briefly discuss the GM model at the tree level in Section~\ref{sec:model}. Section~\ref{sec:renorm} provides a detailed prescription for renormalizing the model at the one-loop level. We present the one-loop result for the $\rho_0$ parameter within the tadpole-free $\msbar$ scheme in Section~\ref{sec:rho-NS}. We discuss the dependence of $\rho_0$ on the model parameters in the GM model in Section~\ref{sec:results}. Finally, we conclude in Section~\ref{sec:conclusion}. The one-loop results for self-energies and mixings are listed in the appendices.
\section{The Model}\label{sec:model}
The GM model~\cite{Georgi:1985nv,Chanowitz:1985ug} is a triplet scalar extension of the SM, which preserves custodial $SU(2)_V$ symmetry at the tree level. The Higgs sector of the GM model contains one scalar doublet $(\phi)$ with hypercharge $Y_\phi=1/2$, one real triplet scalar $(\xi)$ with hypercharge $Y_\xi=0$, and one complex triplet scalar $(\chi)$ with hypercharge $Y_\chi=1$. At tree-level, the Higgs Lagrangian is given by
\begin{equation}
\mathcal{L}_{H}= \frac{1}{2}\mathrm{Tr}[(D_\mu \Phi)^\dagger(D^\mu \Phi)]+\frac{1}{2}\mathrm{Tr}[(D_\mu \Delta)^\dagger(D^\mu \Delta)]-V_{\text{tree}}(\Phi,\Delta)\,,
    \label{eq:Lag-H}
\end{equation}
which is invariant under the global $SU(2)_L\times SU(2)_R$ symmetry. Following this symmetry, the doublet $\phi=(\phi^+ \;\phi^0)^T$, and the two triplets  $\xi=(\xi^+ \;\xi^0\;-\xi^{+*})^T$ and $\chi=(\chi^{++} \;\chi^+ \;\chi^0)^T$ can be written as 
\begin{equation}
\Phi=
\begin{pmatrix}
\phi^{0 *} & \phi^+ \\
-\phi^{+*} & \phi^0 
\end{pmatrix}, \quad \text{and}\quad  
\Delta= 
\begin{pmatrix}
\chi^{0 *} & \xi^+ & \chi^{++} \\
-\chi^{+*} & \xi^0 & \chi^+ \\
\chi^{++*} & -\xi^{+*} & \chi^0
\end{pmatrix}.
\label{GM_fields}
\end{equation}
The covariant derivatives for these scalar fields are defined as 
\begin{align}
    D_\mu\Phi&=\partial_\mu\Phi-igW_\mu^a\tau^a\Phi +ig'B_\mu\tau^3\Phi\,,\\
    D_\mu\Delta&=\partial_\mu\Delta-igW_\mu^a t^a\Delta +ig'B_\mu t^3\Delta\,,
\end{align}
where $\tau^a$ and $t^a$ are the two and three dimensional representations of the $SU(2)$ generators, respectively. The most general scalar potential invariant under the global $SU(2)_L\times SU(2)_R$ symmetry, in the notation of~\cite{Chiang:2015amq}, is given by\footnote{In this work, we consider a CP-conserving GM potential assuming all parameters are real, and we use $\tilde{\mu}_1$, $\tilde{\mu}_2$ instead of $\mu_1$, $\mu_2$ as in Ref.~\cite{Chiang:2015amq}. Note that explicit CP violation is not allowed (at tree-level) in the GM model due to global $SU(2)_L\times SU(2)_R$ symmetry.}
\begin{eqnarray}
\nonumber
V_{\text{tree}}(\Phi, \Delta) &=& 
\frac{1}{2}m^2_1 \tr[\Phi^\dagger \Phi]+\frac{1}{2}m^2_2 \tr[\Delta^\dagger\Delta]+\lambda_1(\tr[\Phi^\dagger \Phi])^2+ \lambda_2 (\tr[\Delta^\dagger \Delta])^2\nonumber \\
&&+ \lambda_3 \tr[(\Delta^\dagger \Delta)^2]+\lambda_4 \tr[\Phi^\dagger \Phi] \tr[\Delta^\dagger \Delta]+\lambda_5 \tr[\Phi^\dagger \tau^a \Phi \tau^b] \tr[\Delta^\dagger t^a \Delta t^b] \nonumber\\
&&+\tilde{\mu}_1 \tr[\Phi^\dagger \tau^a \Phi \tau^b](P^\dagger \Delta P)_{ab} + \tilde{\mu}_2 \tr[\Delta^\dagger t^a \Delta t^b](P^\dagger \Delta P)_{ab}\,,
\label{GM_pot}
\end{eqnarray}
where the square matrix $P$ diagonalizes the three-dimensional representation of the $SU(2)$ generators, as discussed in~\cite{Chiang:2015amq}, and $a,b=1,2,3$. After the EWSB, we redefine the neutral components of the fields as,
\begin{equation}
\phi^0\to\frac{v_\phi}{\sqrt{2}}+\frac{1}{\sqrt{2}}\left(\phi^0+i\phi'\right)\,,\quad \chi^0\to v_\chi+\frac{1}{\sqrt{2}}\left(\chi^0+i\chi'\right)\,,\quad \xi^0\to v_\xi+\xi^0\,,
\end{equation}
with $v_\chi=v_\xi\equiv v_\Delta$ to ensure that the potential remains invariant under custodial $SU(2)_V$ symmetry. The Goldstone bosons,
\begin{eqnarray}
G^+=c_\beta \phi^+ + s_\beta \frac{1}{\sqrt{2}}(\chi^++\xi^+)\,,\quad \text{and}\quad  G^0=c_\beta \phi'+s_\beta \chi'\,,
\label{GM:goldstones}
\end{eqnarray}
show up in the longitudinal mode of massive $W^+$ and $Z^0$ gauge bosons. Here, we have used the notation $s_\beta=2\sqrt{2} v_\Delta/v$ and $c_\beta=v_\phi/v$
with $v^2 =v_\phi^2+8v_\Delta^2$. The gauge boson masses are obtained to be $m_W^2=m_Z^2 \cos^2\theta_W=\frac{1}{4}g^2 v^2$, where $\theta_W$ is the weak mixing angle.  Hence, $\rho\equiv m_W^2\sec^2\theta_W/m_Z^2=1$ at tree level. The mass eigenstates of the physical scalars are expressed as follows:
\begin{align}
H_5^{++}&=\chi^{++} \,,&\quad H_5^+&=\frac{1}{\sqrt{2}}(\chi^+-\xi^+) \,, &\quad H_5^0&=  \frac{1}{\sqrt{3}}\left(\sqrt{2}\xi^0- \chi^0\right)\,,\\
\nonumber
& & H_3^+&=-s_\beta \phi^+ + c_\beta \frac{1}{\sqrt{2}}(\chi^++\xi^+) \,,& \quad H_3^0&=-s_\beta \phi'+ c_\beta \chi'\,, \\
\nonumber
& & h&=c_\alpha \phi^0 - s_\alpha \frac{1}{\sqrt{3}}\left(\xi^0+\sqrt{2} \chi^0\right)\,,&\quad
H_1&=s_\alpha \phi^0 + c_\alpha \frac{1}{\sqrt{3}}\left(\xi^0+\sqrt{2} \chi^0\right)\,,
\label{GM:physical_states}
\end{align}
with $m_{H_1}>m_h$. Here, $\alpha$ denotes the mixing angle between the custodial singlet CP-even scalars $\phi^0$ and $(\xi^0+\sqrt{2} \chi^0)/\sqrt{3}$. Due to the global $SU(2)_R$ symmetry of the scalar potential, the  states $\left\{H_5^{\pm \pm}, H_5^{\pm}, H_5^0 \right\}$ form a custodial fiveplet, $\{H_3^{\pm}, H_3^0\}$ form a custodial triplet, $h$ and $H_1$ are two custodial singlets in the mass-eigenstate basis. The mass eigenstates have definite custodial $SU(2)_V$ quantum numbers, and the scalars with different custodial $SU(2)_V$ quantum numbers do not mix. The lightest custodial singlet $h$ is identified as the observed SM-like Higgs boson. The members of each custodial multiplet are mass-degenerate, which are denoted by~\cite{Chiang:2015amq} 
\begin{align*}
m_{H_5^{++}}^2=m_{H_5^+}^2=m_{H_5^0}^2&\equiv m_{H_5}^2\,,\nonumber\\
    m_{H_3^{+}}^2=m_{H_3^0}^2&\equiv m_{H_3}^2\,.
\end{align*}

\section{One-loop renormalization of the GM model}\label{sec:one-loop}
For the one-loop computation, we use the $\msbar$ scheme~\cite{Marciano:1979yg,Antonelli:1980zt,Fanchiotti:1989wv,Sirlin:1989uf, Passarino:1989ey,Passarino:1990xx,Fanchiotti:1992tu}, unless stated otherwise.\footnote{One of the advantages of choosing this scheme is that it uses the Lagrangian parameters as input, where the presence of large loop corrections is transparent. Also, note that on-shell renormalization~\cite{PhysRevD.22.971,Bohm:1986rj,Hollik:1988ii} of a model is not possible if it does not have a sufficiently large number of free parameters at the tree level to absorb all quantum corrections at one-loop. This is the case for the GM model~\cite{Braathen:2017izn}.} In this scheme, we prescribe the counter-terms and fix the $\msbar$  quantities such as $\alpha_e\equiv\alpha_e(m_Z)$, $ s_{\theta}^2\equiv\sin^2\theta_W(m_Z)$ from low-energy experiments, to describe the physics at the $m_Z$ scale. In the SM and its scalar extensions (singlet, doublet, triplet, etc.), these renormalized quantities are functions of scalar masses, and these scalar masses can be constrained through their effect on quantum corrections to these parameters. The one-loop calculations will be performed in the t'Hooft-Feynman gauge ($\xi=1$). All the one-loop amplitudes are computed using \texttt{FeynRules-2.3}~\cite{Christensen:2008py} and \texttt{FeynArts-3.11}~\cite{Hahn:2000kx}.

\subsection{Renormalization}\label{sec:renorm}
\begin{figure}[htb!]
    \centering
    \includegraphics[width=\textwidth]{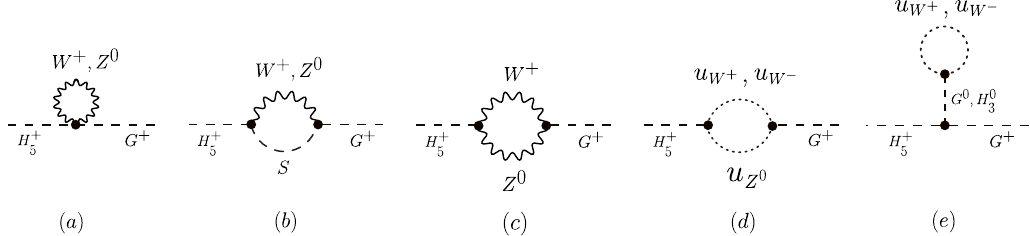}
    \caption{Feynman diagrams for the $H_5^+$-$G^+$ mixing, where the scalars $S=\{G^+,G^0,H_5^0,H_5^+,H_5^{++}\}$.}
    \label{fig:H5pGpMix}
\end{figure}

Here, we discuss the renormalization of the GM model in detail. The GM model has a global $ SU(2)_L \times SU(2)_R$ symmetry at the tree level. Owing to this symmetry, the mixings $H_5^+ \text{--} G^+$,  $H_5^+ \text{--} W^+$, $H_5^+ \text{--} H_3^+$, $H_5^0 \text{--} h$, and $H_5^0 \text{--} H_1$ are absent at tree level. However, at the one-loop level, these mixings are induced and generate divergent contributions~\cite{Gunion:1990dt}. The Ward identity imposes a relation between the first two mixings, $H_5^+ \text{--} G^+$ and $H_5^+ \text{--} W^+$ (evaluated at $p^2 = m_{H_5}^2$),
\begin{equation}\label{WI:H5W}
    m_{H_5}^2\Sigma_{H_5^+W^+}^\mu=p^\mu m_W  \Sigma_{H_5^+G^+}\,,
\end{equation}
where $p^\mu$ is the incoming momentum of $H_5^+$. The relevant part of the bare Lagrangian for $H_5^+ \text{--} G^+$ mixing is\footnote{Throughout this paper, we denote bare parameters with a superscript ``0" (for example, $v_\chi^0$ and $v_\xi^0$ for triplet VEVs).}
\begin{equation}\label{eq:BareLag-charged}
    \mathcal{L}_{H}\supset -\frac{v_\chi^0}{v^0}\Big[\sqrt{2}\Big(T_\chi^0-T_\xi^0\Big)-\Big(v_\chi^0-v_\xi^0\Big)\Big\{\Lambda_1^0+\sqrt{2}\Xi_1^0\Big\}\Big]H_5^+G^-\,,
\end{equation}
where 
\begin{align}\label{eq:charged-notation}
  \Xi_1^0&=\Big(v_\chi^0+v_\xi^0\Big)\frac{\mu_1^0}{v_\xi^0}+\frac{\mu_2^0}{v_\xi^0}\frac{(v_\phi^0)^2}{4v_\chi^0}\,,\nonumber\\
    \Lambda_1^0&=\sqrt{2}\kappa_1^0 \Big(v_\chi^0+v_\xi^0\Big)-\frac{1}{2}\kappa_3^0 \frac{2v_\chi^0+v_\xi^0}{v_\xi^0}\frac{(v_\phi^0)^2}{v_\chi^0}\,,
\end{align}  
and the tree-level expressions of $T_\phi^0,T_\xi^0$, and $T_\chi^0$ are given in Eq.~(\ref{eq:tadpole-tree}). At the tree level, the right-hand side (RHS) of Eq.~(\ref{eq:BareLag-charged}) is zero due to the extremum conditions given in Eq.~(\ref{eq:tadpole-tree}) together with the relation $v_\chi=v_\xi$. However, these conditions no longer hold once loop contributions are taken into account. As a result, a nonzero $H_5^+\text{--}G^+$ mixing is generated at one-loop (corresponding Feynman diagrams are shown in Figure~\ref{fig:H5pGpMix}).
 The sum of all tadpole contributions (shown in Figure~\ref{fig:H5pGpMix}(e)) is zero because the couplings satify $g_{s_iu_{W^+}\bar{u}_{W^+}}=-g_{s_iu_{W^-}\bar{u}_{W^-}}$ for $s_i=G^0$ or $H_3^0$, where $u_{W^+}$ and $u_{W^-}$ are the Faddeev–Popov (FP) ghost fields corresponding to the $W^+$ and $W^-$ gauge bosons, respectively. For $s_i=G^0$ or $H_3^0$, all trilinear couplings $s_is_ks_k$ ($s_k=h,H_1^0,G^+,G^0,H_3^0,H_3^+,H_5^0,H_5^+,$ or $H_5^{++}$) and $s_iVV$ ($V=W^+, Z^0$) are zero in the  $SU(2)_R$-symmetric limit. Furthermore, scalar-loop diagrams of classes (a) and (b) are absent due to $SU(2)_L\times SU(2)_R$ symmetry in the scalar potential. Explicit one-loop expressions for $H_5^+\text{--}G^+$ mixing are given in Appendix~\ref{App:results}. The net $1/\hat\epsilon$ infinite term of $H_5^+\text{--}G^+$ mixing is  
\begin{equation}\label{eq:mixH5pGp}
    \Sigma_{H_5^+G^+}(p^2)=\frac{1}{\hat\epsilon}\frac{g'^2s_\beta}{64\pi^2}\Big[6m_W^2\Big(2+\tan^2\theta_W\Big)-m_{H_5}^2-4p^2\Big]\,.
\end{equation}
The $1/\hat\epsilon$ infinite term for the $H_5^+\text{--}G^+$ mixing is proportional to the $U(1)_Y$ coupling constant $ g'$. Hence, $\Sigma_{H_5^+G^+}$ is a finite quantity in the limit $ g'\to 0$. Note that hypercharge gauge interactions with the coupling constant $g'$ break the $SU(2)_R$ symmetry.  Following~\cite{Gunion:1990dt}, we exactly cancel the mixing $H_5^+\text{--}G^+$ on the $H_5$ mass shell by subtracting $H_5^+\text{--}G^+$ mixing at $p^2=m_{H_5}^2$,
\begin{align}\label{eq:H5Gmix_tilde}
    \widetilde{\Sigma}_{H_5^+G^+}(p^2)&= \Sigma_{H_5^+G^+}(p^2) +(p^2-m_{H_5}^2)s_\beta\delta \eta - \Sigma_{H_5^+G^+}(m_{H_5}^2)\,, 
\end{align}
where the counter-term $\delta \eta=g'^2/(16\pi^2\hat\epsilon)$ is necessary to cancel the $p^2/\hat\epsilon$ infinity. This is a consequence of the fact that the wavefunction renormalizations of the $\chi^+$ and $\xi^+$ fields are different (recall that $H_5^+\propto (\chi^+-\xi^+)$) due to $SU(2)_R$ breaking at the one loop. 
The renormalized quantity $\widetilde{\Sigma}_{H_5^+G^+}$ is free from ultraviolet divergence. Due to the Ward identity in Eq.~(\ref{WI:H5W}), the one-loop $H_5^+ \text{--} W^+$ mixing is also finite and vanishes on the $H_5$ mass shell, ensuring consistency with gauge symmetry.
\begin{figure}[htb!]
    \centering
    \includegraphics[width=0.9\textwidth]{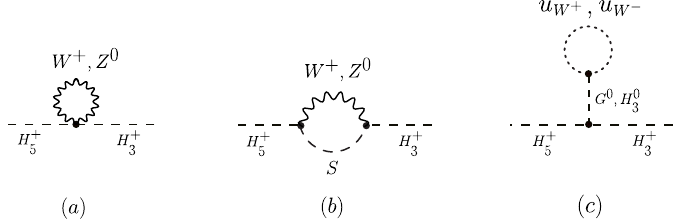}
    \caption{Feynman diagrams for the $H_5^+$-$H_3^+$ mixing, where the scalars $S=\{H_3^+,H_3^0,H_5^0,H_5^+,H_5^{++}\}$.}
    \label{fig:fd-H5pH3p}
\end{figure}

At one loop, the mixing of the singly charged scalars $H_5^+$ and $H_3^+$ is shown in Figure~\ref{fig:fd-H5pH3p}. As discussed above, the sum of all tadpole contributions (shown in Figure~\ref{fig:fd-H5pH3p}(c)) is zero. For $H_5^+ \text{--} H_3^+$ mixing, diagrams in class (b) involving only gauge bosons or Faddeev–Popov ghosts are absent, since $g_{H_3^-W^+Z}=0$ in the  $SU(2)_R$-symmetric limit. Additionally, diagrams in classes (a) and (b) with scalar loops are absent due to $SU(2)_L\times SU(2)_R$ symmetry in the scalar potential. The complete expression for the mixing of $H_5^+\text{--}H_3^+$ is given in Appendix~\ref{App:results}. The relevant part of the bare Lagrangian for $H_5^+\text{--}H_3^+$ mixing is  
\begin{align}\label{eq:BareLagH5pH3p}
    \mathcal{L}_{H}\supset -\frac{v_\phi^0}{2\sqrt{2}v^0}\Big[\sqrt{2}\Big(T_\chi^0-T_\xi^0\Big)&-\Big(v_\chi^0-v_\xi^0\Big)\Big\{\Lambda_1^0+\sqrt{2}\Xi_1^0\Big\}\nonumber\\
    &+ \frac{(v^0)^2}{2\sqrt{2}v_\chi^0}\Big\{\sqrt{2}\mu_3^0-\mu_2^0+v_\chi^0\Big(\sqrt{2}\kappa_3^0-\kappa_2^0\Big)\Big\}\Big]H_5^+H_3^-\,,
\end{align}
where $\Xi_1^0$ and $\Lambda_1^0$ are given in Eq.~(\ref{eq:charged-notation}). The net $1/\hat\epsilon$ infinite term of this mixing is given by
\begin{equation}\label{eq:mixH5pH3p}
    \Sigma_{H_5^+H_3^+}(p^2)=-\frac{1}{\hat\epsilon}\frac{g'^2c_\beta}{64\pi^2}\Big[m_{H_5}^2+m_{H_3}^2-6m_W^2\tan^2\theta_W+4p^2\Big]\,.
\end{equation}
Because of the presence of the same bare terms as in Eqs.~(\ref{eq:BareLag-charged}) and (\ref{eq:BareLagH5pH3p}), the $H_5^+\text{--}H_3^+$ mixing can be written as, 
\begin{align}\label{eq:H5pH3p-bar}
    \bar{\Sigma}_{H_5^+H_3^+}(p^2)&= \Sigma_{H_5^+H_3^+}(p^2) +(p^2-m_{H_5}^2)c_\beta\delta \eta - \frac{c_\beta}{s_\beta}\Sigma_{H_5^+G^+}(m_{H_5^2})\,.
\end{align}
The $p^2/\hat\epsilon$ term in the $H_5^+ \text{--} H_3^+$ mixing is also cancelled by the same counter-term $\delta\eta$ used in Eq.~(\ref{eq:H5Gmix_tilde}).
From Eqs.~(\ref{eq:mixH5pH3p}) and (\ref{eq:H5pH3p-bar}), we observed that the modified mixing $\bar{\Sigma}_{H_5^+H_3^+}$ is not an ultraviolet finite quantity. To renormalize the mixing $\bar{\Sigma}_{H_5^+H_3^+}$ at the one-loop, we add $SU(2)_R$-breaking counter-terms. The most general scalar potential invariant under the local $SU(2)_L\times U(1)_Y$ is given in Appendix~\ref{App:MostGenPotential}. 
The renormalized parameters $X$ are defined in terms of the bare ones $X^0$ through $X^0\to X+\delta X$, where $\delta X$ denotes the corresponding counter-term. The most general counter-term potential invariant under local $SU(2)_L\times U(1)_Y$ symmetry can be written as
\begin{align}
V_{\text{CT}}(\phi,\chi,\xi)&=-\delta m_\phi^2\big(\phi^\dagger\phi\big)-\delta m_\xi^2\big(\xi^\dagger\xi\big)-\delta m_\chi^2\big(\chi^\dagger\chi\big)+\delta \mu_1\big(\chi^\dagger t^a\chi\big)\xi^a+\delta \mu_2\big(\phi^\dagger \tau^a\phi\big)\xi^a\nonumber\\
&\,\quad+\delta \mu_3\Big[\big(\phi^T\epsilon\tau^a\phi\big)\tilde{\chi}^a+\text{h.c.}\Big]
+\delta \lambda_\phi\big(\phi^\dagger\phi\big)^2+\delta \lambda_\xi\big(\xi^\dagger\xi\big)^2+\delta \lambda_\chi\big(\chi^\dagger\chi\big)^2\nonumber\\
&\,\quad+\delta\tilde{\lambda}_\chi \big|\tilde{\chi}^\dagger\chi \big|^2+\delta\lambda_{\phi\xi}\big(\phi^\dagger\phi\big)\big(\xi^\dagger\xi\big)
+\delta\lambda_{\phi\chi}\big(\phi^\dagger\phi\big)\big(\chi^\dagger\chi\big)+\delta\lambda_{\chi\xi}\big(\chi^\dagger\chi\big)\big(\xi^\dagger\xi\big)\nonumber\\
&\,\quad+\delta\kappa_1\big|\xi^\dagger\chi \big|^2+\delta\kappa_2\big(\phi^\dagger\tau^a\phi\big)\big(\chi^\dagger t^a\chi\big)+\delta\kappa_3\Big[\big(\phi^T\epsilon\tau^a\phi\big)\big(\chi^\dagger t^a\xi\big)+\text{h.c.}\Big]\,,
\label{eq:GM-CTpot}
\end{align} 
where $\tilde{\chi}=(\chi^{0*} \;-\chi^{+*} \;\chi^{++*})^T$ is the charge conjugate of the complex triplet field $\chi$. In order to exactly cancel the mixing $\bar{\Sigma}_{H_5^+H_3^+}$ on the $H_5$ mass shell, we fix (at one-loop order)
\begin{equation}
    \sqrt{2}\delta\kappa_3-\delta \kappa_2+\frac{1}{v_\chi}\Big(\sqrt{2}\delta\mu_3-\delta\mu_2\Big)=\frac{8}{v^2c_\beta}\bar{\Sigma}_{H_5^+H_3^+}(m_{H_5}^2)\,.
\end{equation}
Note that when the $SU(2)_R$ symmetry is exact, the counter-terms also satisfy $\sqrt{2}\delta\kappa_3=\delta \kappa_2$ and $\sqrt{2}\delta\mu_3=\delta\mu_2$ (see Appendix~\ref{App:MostGenPotential}). Hence, the renormalized quantity for $H_5^+ \text{--} H_3^+$ mixing is 
\begin{align}\label{eq:H5pH3p-tilde}
    \widetilde{\Sigma}_{H_5^+H_3^+}(p^2)&= \bar{\Sigma}_{H_5^+H_3^+}(p^2) - \bar{\Sigma}_{H_5^+H_3^+}(m_{H_5}^2)\,.
\end{align}
As a result, the custodial $SU(2)_R$ eigenstate $H_5^+$ remains well-defined at its mass shell $p^2=m_{H_5}^2$, and the on-shell decay of $H_5^+$ into fermions via mixing with $H_3^+$ vanishes. In the GM model, $H_3^+$ is the only singly-charged scalar that couples to fermions. The Yukawa interactions involving lepton doublets and the Higgs triplets are omitted, given our assumption that $v_\Delta$ is $\mathcal{O}(1)$ in this study~\cite{Chiang:2012cn}.

For the $H_5^0$ mixing at the one-loop, we choose the custodial singlets $\phi^0$ and $H_1'=(\xi^0+\sqrt{2}\chi^0)/\sqrt{3}$ and work in the guage basis.\footnote{Here, we work in the gauge basis solely to simplify the calculations. Extending the analysis to the mass-eigenstate basis is straightforward, with the physical scalars $h=c_\alpha \phi^0-s_\alpha H_1'$ and $H_1=s_\alpha \phi^0+c_\alpha H_1'$.} The relevant part of the bare Lagrangian that contributes to the $H_5^0 \text{--} \phi^0$ mixing is
\begin{align}\label{eq:BareLag-H50h}
    \mathcal{L}_{H}\supset -\frac{v_\phi^0}{4\sqrt{3}}\Big[\sqrt{2}v_\chi^0\Big(\kappa_2^0-\sqrt{2}\kappa_3^0+2\lambda_{\phi\chi}^0&-4\lambda_{\phi\xi}^0\Big)+\sqrt{2}\Big(\mu_2^0-\sqrt{2}\mu_3^0\Big)\nonumber\\
    &-\Big(v_\chi^0-v_\xi^0\Big)\Big\{2\kappa_3^0-4\sqrt{2}\lambda_{\phi\xi}^0\Big\}\Big]H_5^0\phi^0\,.
\end{align}
At the tree level, the RHS of Eq.~(\ref{eq:BareLag-H50h}) vanishes due to the relations given in Eq.~(\ref{eq:GMconstraints}) and the condition $v_\chi=v_\xi$. However, at the one-loop, the RHS of Eq.~(\ref{eq:BareLag-H50h}) is not zero. We exactly cancel the mixing $H_5^0 \text{--} \phi^0$ on the $H_5$ mass shell by subtracting the mixing at $p^2=m_{H_5}^2$,
\begin{align}\label{eq:H50hmix_tilde}
    \widetilde{\Sigma}_{H_5^0\phi^0}(p^2)&= \Sigma_{H_5^0\phi^0}(p^2)  - \Sigma_{H_5^0\phi^0}(m_{H_5}^2)\,.
\end{align}
Similarly, the bare Lagrangian for $H_5^0 \text{--} H_1'$ mixing is given by
\begin{align}\label{eq:BareLag-H50H1p}
    \mathcal{L}_{H}\supset -\frac{\sqrt{2}}{3}\Big[\Big(T_\chi^0-T_\xi^0\Big)&+\Big(v_\chi^0-v_\xi^0\Big)\Big\{\Lambda_2^0-\sqrt{2}\Xi_2^0\Big\}\nonumber\\
    &+ (v_\chi^0)^2\Big(2\lambda_\chi^0-4\lambda_\xi^0-\lambda_{\chi\xi}^0\Big)-\frac{(v_\phi^0)^2}{8v_\xi^0}\Big(\mu_2-\sqrt{2}\mu_3\Big)\Big]H_5^0H_1'\,,
\end{align}
where 
\begin{align}\label{eq:neutral-notation}
  \Xi_2^0&=\frac{v_\chi^0}{2\sqrt{2}}\frac{\mu_1^0}{v_\xi^0}+\frac{\mu_3^0}{v_\chi^0}\frac{(v_\phi^0)^2}{8v_\xi^0}\,,\nonumber\\
    \Lambda_2^0&=4\lambda_\xi^0 \Big(v_\chi^0+v_\xi^0\Big)+v_\chi^0\lambda_{\chi\xi}^0+\kappa_3^0 \frac{2v_\chi^0+v_\xi^0}{v_\xi^0}\frac{(v_\phi^0)^2}{4\sqrt{2}v_\chi^0}\,,
\end{align}
and the tree-level expressions of $T_\xi^0$ and $T_\chi^0$ are given in Eq.~(\ref{eq:tadpole-tree}). The RHS of Eq.~(\ref{eq:BareLag-H50H1p}) is not zero beyond the tree level. The renormalized quantity for the mixing $H_5^0 \text{--} H_1'$ is given by 
\begin{align}\label{eq:H50H1p-tilde}
    \widetilde{\Sigma}_{H_5^0H_1'}(p^2)&= \Sigma_{H_5^0H_1'}(p^2) -\frac{2\sqrt{2}}{3}(p^2-m_{H_5}^2)\delta \eta - \Sigma_{H_5^0H_1'}(m_{H_5^2})\,,
\end{align}
where the counter-term $\delta \eta$ is given in Eq.~(\ref{eq:H5Gmix_tilde}). As a result, the loop–induced mixing does not generate any on-shell decay of $H_5^0$ to fermions. In the spirit of $SU(2)_R$-symmetric theory, we renormalized one-loop mixings on the $H_5$ mass shell where the $SU(2)_R$ symmetry is maximal.

The one-loop corrections to the remaining tree-level relations (originating from the kinetic part of the Higgs Lagrangian) are finite since the theory is renormalizable. In the mass-eigenstate basis, the one-loop correction to the tree-level relation $m_W^2=c_\theta^2m_Z^2$ originated from the gauge-boson self-energies is given (at $p^2=0$) by~\cite{Peskin:1990zt,Peskin:1991sw}
\begin{equation}\label{eq:DeltaRho}
    \Delta\rho=\frac{1}{16\pi^2m_W^2}\text{Re}\Big[\Sigma_{WW}(0)-c_\theta^2\Sigma_{ZZ}(0)-2s_\theta c_\theta\Sigma_{\gamma Z}(0)-s_\theta^2 \Sigma_{\gamma\gamma}(0) \Big]\,,
\end{equation}
where the notation $s_\theta$ and $c_\theta$ are the sine and cosine of the weak mixing angle ($\theta_W$), respectively. $\Delta \rho$ receives contributions from both the SM and possible new physics effects. To constrain the new physics contribution in the GM model, we define~\cite{ParticleDataGroup:2016lqr}
\begin{equation}
    \rho_0\equiv \frac{m_W^2}{m_Z^2c_\theta^2\rho}\,,
\end{equation}
where $\rho$ is computed in the SM limit. Here, $\rho_0=1$ at the tree-level. The non-standard (NS) one-loop contribution to $\rho_0$ is equivalent to the Peskin-Takeuchi $T$ parameter~\cite{Peskin:1990zt,Peskin:1991sw}, i.e.,
\begin{equation}\label{eq:T-para}
    \rho_0-1=\Delta\rho^{\text{NS}}\simeq \alpha_e T\,.
\end{equation}
 After introducing the three renormalization conditions (for $\alpha_e, G_\mu,$ and $m_Z^2$) which fix the counter-terms at one-loop, 
\begin{align}
\frac{\delta \alpha_e}{\alpha_e}&=\frac{d}{dp^2}\Sigma_{\gamma\gamma}(p^2)\Big|_{p^2=0}+\frac{2s_\theta}{c_\theta}\frac{\Sigma_{\gamma Z}(0)}{m_Z^2}\,,\\
\frac{\delta G_\mu}{G_\mu}&=\frac{\delta \alpha_e}{\alpha_e}-\frac{\delta m_W^2}{m_W^2}-\frac{\delta s_\theta^2}{s_\theta^2}\,,\\
    \delta m_Z^2&=\text{Re}\:\Sigma_{ZZ}(m_Z^2)\,,
\end{align}
with $\delta m_W^2=\text{Re}\:\Sigma_{WW}(m_W^2)$\,, and 
\begin{align}
    \frac{\delta s_\theta^2}{s_\theta^2}&=\frac{c_\theta^2}{s_\theta^2}\left(\frac{\delta m_Z^2}{m_Z^2}-\frac{\delta m_W^2}{m_W^2}\right)\,,
\end{align}
we find that the quantity $\Delta\rho^{\text{NS}}$ (or $T$ parameter)  is ultraviolet divergent. Although the divergences in the $\Delta\rho^{\text{NS}}$ and the mixing $H_5^+ \text{--} W^+$ originate from the same custodial $SU(2)_V$-breaking term proportional to $(v_\chi-v_\xi)$, we find that the one-loop corrections and the $1/\hat\epsilon$ infinite terms of these two quantities are not proportional. The relevant term in the bare Higgs Lagrangian is 
\begin{equation}\label{eq:bareMW}
    \mathcal{L}_H\supset \frac{1}{4}(g^0)^2\Big[(v^0)^2-4\Big\{(v_\chi^0)^2-(v_\xi^0)^2\Big\}\Big]W_\mu^+W^{-\mu} -\frac{ig^0}{\sqrt{2}}\Big(v_\chi^0-v_\xi^0\Big)\Big[W_\mu^-\partial^\mu H_5^+ +\text{h.c.}\Big]\,,
\end{equation}
and the corresponding two-point functions have the following  Lorentz structure:
\begin{align}\label{eq:H5W-2p}
    \Sigma_{H_5^+W^+}^\mu(p)&=p^\mu\Sigma_{H_5^+W^+}(p^2),\\
    \Sigma_{W^+W^+}^{\mu\nu}(p)&=g^{\mu\nu}\Sigma_{W^+W^+}^{T}(p^2)+\frac{p^\mu p^\nu}{p^2}\Big[\Sigma_{W^+W^+}^{L}(p^2)-\Sigma_{W^+W^+}^{T}(p^2)\Big]\,,\label{eq:WW-2p}
\end{align}
where we denote the longitudinal and transverse parts of the self-energy with superscripts ``L" and ``T", respectively. The $g^{\mu\nu}$ term in Eq.~(\ref{eq:WW-2p}) contributes to the $\Delta\rho$ (see Eq.~(\ref{eq:DeltaRho}) where $ \Sigma_{W^+W^+}^{T}\equiv \Sigma_{WW}$). On the other hand, the $p^{\mu}$ term in Eq.~(\ref{eq:H5W-2p}) is related to the Higgs-Goldstone mixing $H_5^+ \text{--} G^+$ (at $p^2=m_{H_5}^2$) through the Ward identity (see Eq.~(\ref{WI:H5W})). However, we find that their tadpole contributions remain proportional, in agreement with Ref.~\cite{Gunion:1990dt}. This is because the tadpole effects arise only through the VEV counter-terms.

Here, in the GM model, $\delta s_\theta^2$ is an independent counter-term, and an extra renormalization condition is required to fix it. We can choose $\rho_0$ (or $s_\theta^2$, or $m_W$) as an input parameter and fix the counter-term,
\begin{equation}\label{eq:delta_stheta}
    \frac{\delta s_\theta^2}{ s_\theta^2}=\frac{c_\theta^2}{s_\theta^2}\left(\frac{\delta m_Z^2}{m_Z^2}-\frac{\delta m_W^2}{m_W^2}+\frac{\delta \rho_0}{\rho_0}\right)\,,
\end{equation}
where $\delta \rho_0$ is the counter-term for the parameter $\rho_0$. Eq.~(\ref{eq:delta_stheta}) holds for any renormalizable scalar extension of the SM that preserves custodial symmetry. For example, in the case of singlet and doublet extensions, $\delta \rho_0$ is equal to zero, and the prescription for renormalization is analogous to that of the SM. In contrast, for the GM model, even though the gauge boson -- fermion sector is characterized by three free parameters at tree level, an additional input is needed at one loop to fully determine the electroweak precision observables.

\subsection{Corrections to the $\boldsymbol{\rho}$ parameter}\label{sec:rho-NS}
\begin{figure}[htb!]
    \centering
    \includegraphics[width=0.8\textwidth]{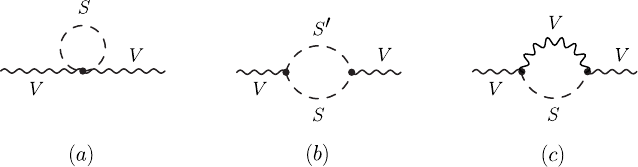}
    \caption{One-loop Feynman diagrams for the vector boson self-energies involving scalar fields. }
    \label{fig:fd-SVV}
\end{figure}

In the tadpole-free $\msbar$ scheme, tadpole contributions to the self-energies of the gauge bosons are exactly canceled by including the appropriate tadpole counterterms at the one-loop level. For the neutral scalars 
$h$ and $H_1$, the self-energies of the 
$W$ and $Z$ bosons satisfy the custodial relation $\Sigma_{WW}^{\phi_i}=c_\theta^2\Sigma_{ZZ}^{\phi_i}$ for $\phi_i=h,H_1$ (i.e.,  $g_{\phi_iW^+W^-}= c_\theta^2 g_{\phi_iZZ}$), which ensures that the tadpole contributions of  $h$ and $H_1$ to $\Delta\rho^{\text{NS}}$ or $T$ parameter (given in Eq.~(\ref{eq:T-para})) vanish individually. In contrast, for $\phi_i=H_5^0$, the tadpole contributions to  $\Delta\rho^{\text{NS}}$ (or $T$ parameter) contain quadratic divergences. This arises from the  coupling relation $g_{H_5^0W^+W^-}=-c_\theta^2( g_{H_5^0ZZ}/2)$. A discussion of the tadpole renormalization at the one-loop level is presented in Appendix~\ref{App:tadpole}. We now turn to the remaining one-particle irreducible (1PI) Feynman diagrams involving scalar fields in the GM model. There are three types of such diagrams, shown in Figure~\ref{fig:fd-SVV}, which contribute to the self-energy of the gauge bosons. 
The one-loop expressions given in Appendix~\ref{App:results} correspond to the sum over all possible configurations of scalar fields, $S$ and $S'$.
In the GM model, physical scalar loops corresponding to diagrams (a) and (b) in the Figure~\ref{fig:fd-SVV} do not contribute to $\Delta \rho^\text{NS}$ due to the global $SU(2)_L\times SU(2)_R$ symmetry of the scalar potential. Consequently, there is no quadratic dependence on the scalar masses in the $\rho$ parameter at one loop.
Considering the remaining non-standard one-loop contribution to $\Delta \rho^\text{NS}$, at the scale $\mu=m_Z$, we find
\begin{align}\label{eq:DeltaRho-results}
    \Delta \rho^\text{NS}&= \frac{g^2}{64\pi^2m_W^2}\Big[(\kappa_V^h)^2\Big\{3F[m_Z^2,m_h^2]-3F[m_W^2,m_h^2]\Big\}+(\kappa_V^{H_1})^2\Big\{3F[m_Z^2,m_{H_1}^2]-3F[m_W^2,m_{H_1}^2]\Big\}\nonumber\\
    &\quad+\frac{1}{3}s_\beta^2\Big\{3F[m_Z^2,m_{H_5}^2]-3F[m_W^2,m_{H_5}^2]\Big\}-3F[m_Z^2,m_h^2]+3F[m_W^2,m_h^2]\nonumber\\
    &\quad +4s_\beta^2m_Z^2s_\theta^2+8s_\beta^2m_W^2\ln c_\theta^2\Big]\,,
\end{align}
where the function $F[m_1^2,m_2^2]$ is defined in Eq.~(\ref{eq:F-func}), and the coupling modifiers $\kappa_V^h$ and $\kappa_V^H$ are given by 
\begin{equation}
    \kappa_V^h=c_\alpha c_\beta-\sqrt{\frac{8}{3}}s_\alpha s_\beta\,,\quad \kappa_V^H=s_\alpha c_\beta+\sqrt{\frac{8}{3}}c_\alpha s_\beta\,.
\end{equation}
The quadratic dependence on the scalar masses in $\Delta \rho^\text{NS}$ (see Eq.~(\ref{eq:DeltaRho-results})) is entirely canceled, and the dependency is only logarithmic. From Eq.~(\ref{eq:DeltaRho-results}), it is also evident that $\Delta \rho^\text{NS}$ depends only logarithmically on the scalar masses $m_{h},m_{H_1},$ and $m_{H_5}$.

\section{Numerical Results}\label{sec:results}
In this section, we present the numerical results of the non-standard one-loop (finite) corrections to the $\rho_0$ parameter in the GM  model. In the $\msbar$ scheme, we use three precise input quantities $\alpha_e, G_\mu$, and $m_Z$ to fix the free parameters in the model at tree-level. The experimental value of these quantities are given by~\cite{ParticleDataGroup:2024cfk} $\alpha_e^{-1}(m_Z)=127.930\pm0.008$, $G_\mu= 1.1663788(6) \times 10^{-5}\;\text{GeV}^{-2}$, and $m_Z= 91.1880 \pm 0.0020$ GeV. To quantify the one-loop effects of the new physics in the $\rho_0$ parameter, we fix the SM-like Higgs mass to its experimental value, $m_h = 125.09$ GeV~\cite{ATLAS:2015yey}. 
\begin{figure}[htb!]
    \centering
    \includegraphics[width=0.85\textwidth]{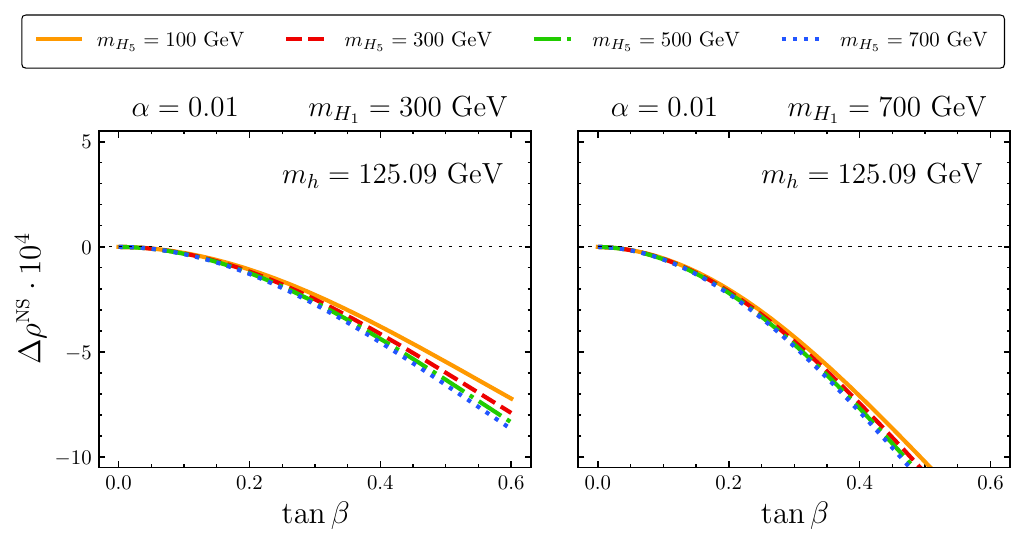}
    \caption{Dependence of $\tan\beta$ on $\Delta\rho^{\text{NS}}$ for various values of $m_{H_1}$ and $m_{H_5}$. The CP-even scalar mixing angle $\alpha$ is fixed at $0.01$. We choose the values of $m_{H_1}$, $m_{H_5}$, and $\alpha$ that are allowed by the theoretical constraints and the latest Run-II Higgs signal strength data from the CMS and
    ATLAS detectors (taken from Ref.~\cite{Chowdhury:2024mfu}).}
    \label{fig:GM_rho_tanb}
\end{figure}
At the tree-level, the $\rho_0$ parameter is independent of the triplet VEV, $v_\Delta$. Beyond the tree-level, the one-loop contribution to $\Delta\rho^{\text{NS}}$ (see Eq.~(\ref{eq:DeltaRho-results})) depends on $v_\Delta$ (or $\tan\beta$) in addition to the non-standard scalar masses $m_{H_1}$, $m_{H_5}$, and the mixing angle ($\alpha$) between the custodial singlet CP-even scalars.

In Figure~\ref{fig:GM_rho_tanb}, we display the variation of $\Delta\rho^{\text{NS}}$ as a function of $\tan\beta$, where we have considered the values of $m_{H_1}$, $m_{H_5}$, and $\alpha$ that are allowed by the theoretical constraints and the latest Run-II Higgs signal strength data from the CMS and ATLAS detectors (see, Ref.~\cite{Chowdhury:2024mfu}). We find that the finite part of
$\Delta\rho^{\text{NS}}$ is negative in the GM model and approaches the Standard Model value (i.e., $\Delta\rho^{\text{NS}}=0$) in the limit of $\tan\beta\to 0$. However, the finite part of the counter-term $\delta\rho_0$ can either be positive or negative. Once the renormalized value of $\rho_0$ (or $s_\theta^2$, or $m_W$) is determined by the experimental inputs, all other electroweak precision observables become predictions of the model. 

\begin{figure}[htb!]
    \centering
    \includegraphics[width=\textwidth]{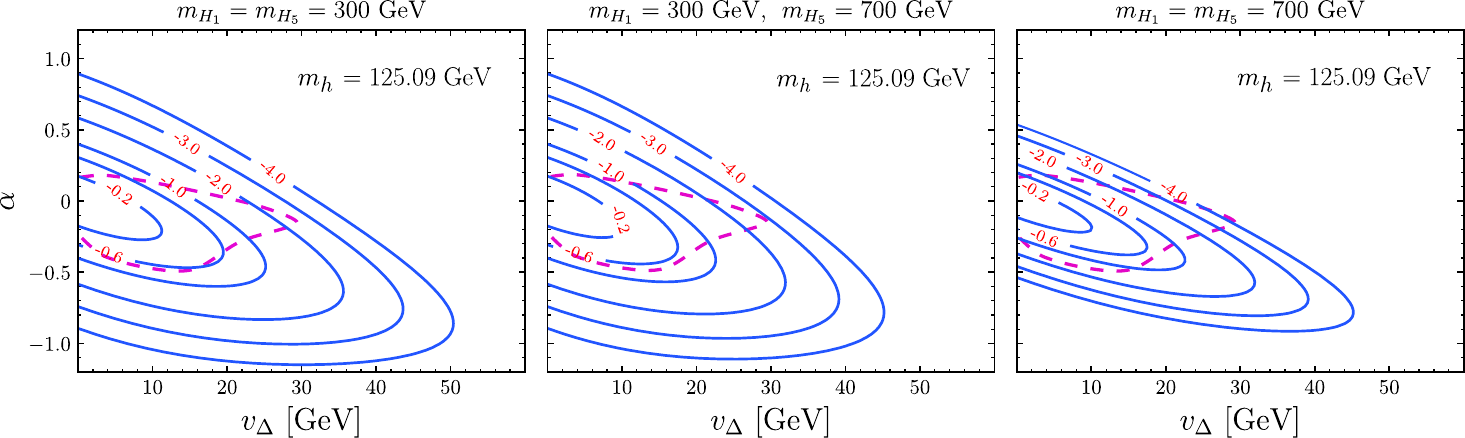}
    \caption{The contours of $\Delta\rho^{\text{NS}}$ (shown in blue color) in the $v_\Delta$ vs. $\alpha$ plane. We denote the value of $\Delta\rho^{\text{NS}}\cdot 10^4$ for each contour in the red color. We choose the values of $m_{H_1}$ and $m_{H_5}$ that are allowed by the theoretical constraints and the latest Run-II Higgs signal strength data from the CMS and ATLAS detectors (taken from Ref.~\cite{Chowdhury:2024mfu}). The magenta dashed contour represents the allowed region from the latest Run-II Higgs signal strength data (taken from Ref.~\cite{Chowdhury:2024mfu}).
    }
    \label{fig:GM_rho_contour}
\end{figure}

To show the dependence of CP-even scalar mixing angle $\alpha$ on $\Delta\rho^{\text{NS}}$, we display the contours of $\Delta\rho^{\text{NS}}$ (shown in blue color) in the $v_\Delta$ vs. $\alpha$ plane of  Figure~\ref{fig:GM_rho_contour}, where the magenta dashed contour represents the allowed region from the latest Run-II Higgs signal strength data (taken from Ref.~\cite{Chowdhury:2024mfu}). We denote the value of $\Delta\rho^{\text{NS}}\cdot 10^4$ for each contour in the red color. We see that the finite part of
$\Delta\rho^{\text{NS}}$ is always negative, and its absolute value increases once we increase the value of $\alpha$ from zero in either positive or negative directions. More to the point, the finite part of
$\Delta\rho^{\text{NS}}$ also depends on the BSM scalar masses $m_{H_1}$ and $m_{H_5}$. In general, the dependence on $m_{H_5}$ and on the difference between $m_{H_5}$ and $m_{H_1}$ is very smooth, as shown in Figures~\ref{fig:GM_rho_tanb} and \ref{fig:GM_rho_contour}. In contrast, if we increase the value of $m_{H_1}$, the $\Delta\rho^{\text{NS}}$ runs into more negative values even in the region where the values of $\alpha$ and $v_\Delta$ are not too large. As mentioned above, the renormalized value of $\rho_0$ is determined from the global EW fit results. The global EW fit yields~\cite{ParticleDataGroup:2024cfk}, $\rho_0= 1.00031\pm 0.00019\,$, which is mostly shifted towards the positive value of $\rho_0-1$. Therefore, to determine whether the constraints on the $v_\Delta$ vs. $\alpha$ plane derived from the latest Run-II Higgs signal strength data remain comparable or become more stringent in the presence of a fourth electroweak input parameter, a full global analysis of the GM model with all four input parameters is required. 
\section{Conclusions}\label{sec:conclusion}
In the GM model, the global $SU(2)_R$ symmetry is not radiatively protected due to hypercharge gauge interactions. As a result, $SU(2)_R$-breaking operators, absent at tree level, arise at one loop and induce ultraviolet divergences. We have calculated one-loop corrections to the $\rho$ parameter in this model. We find that the tadpole contributions to the $\rho$ parameter vanish once the scalar-Goldstone mixing is renormalized. Furthermore, renormalizing the scalar-scalar mixings by introducing $SU(2)_R$-breaking counter-terms in the scalar potential, we show that the $\rho$ parameter at the one-loop is an ultraviolet divergent quantity, when the standard three input ($\alpha_e, G_\mu,$ and $m_Z$) renormalization scheme is employed. The counter-term $\delta s_\theta^2$ appears as an independent quantity, and therefore an additional experimental input is required to fix it. 

Our results show that the one-loop corrections to the $\rho$ parameter depend on $\tan\beta$ and the mixing angle ($\alpha$) between the custodial singlet CP-even scalars, with a mild dependence on the differences between $m_{H_1}$ and $m_{H_5}$. We further find that the one-loop corrections to the $\rho$ parameter are always negative, whereas its corresponding counter-term can take either positive or negative values. Once the renormalized value of the $\rho$ parameter (or $s_\theta^2$, or $m_W^2$) is fixed by experimental input, all remaining electroweak precision observables become predictions of the model.

In summary, this work presents a detailed study of the renormalization prescription for the GM model at the one-loop level, demonstrating that a consistent renormalization requires four independent input parameters to fully specify the electroweak sector. The GM model provides an interesting case study compared to scalar extensions of the SM involving only singlets or doublets. A comprehensive global analysis of this model, incorporating an additional input, lies beyond the scope of the present work and will be studied in the future.

The one-loop renormalization prescription presented here is applicable to other custodial-symmetric Higgs triplet models, such as the extended Georgi-Machacek model~\cite{Kundu:2021pcg,Kundu:2024sip}.

\begin{acknowledgments}
We are thankful to Michael Peskin for helpful discussions and for carefully reading the manuscript and providing valuable comments. We also thank Joydeep Chakrabortty and Kirtimaan Mohan for useful discussions. D.C. and P.M. acknowledge funding from the ANRF, Government of India, under grant ANRF/CRG/2021/007579. P.M. also acknowledges the support from the
Department of Atomic Energy (DAE), Government of India, under Project Identification
Number RTI 4002. A.K. acknowledges ANRF, Government of India, for support through the Project ANRF/CRG/2025/000133.
S.S. acknowledges funding from the MHRD, Government of India, under the Prime Minister’s Research Fellows (PMRF) Scheme. D.C. also acknowledges support from an initiation grant IITK/PHY/2019413 at IIT Kanpur and funding from the Indian Space Research Organisation (ISRO) under grant STC/PHY/2024427Q.
\end{acknowledgments}
\appendix 
\section{One-Loop Scalar Functions }\label{App:PV-func}
Our results for self-energies and mixings at one-loop are given in terms of the following one- and two-point scalar functions,
\begin{align}
    16\pi^2\mu^{2\epsilon}\int \frac{d^Dk}{i(2\pi)^k}\frac{1}{k^2-m^2+i\varepsilon}&=\frac{m^2}{\hat\epsilon}+\bar{A}_0[m^2]\,,\\
     16\pi^2\mu^{2\epsilon}\int \frac{d^Dk}{i(2\pi)^k}\frac{1}{\Big[k^2-m_1^2+i\varepsilon\Big]\Big[(k-p)^2-m_2^2+i\varepsilon\Big]}&=\frac{1}{\hat\epsilon}+\bar{B}_0[p^2,m_1^2,m_2^2]\,,\\
       16\pi^2\mu^{2\epsilon}\int \frac{d^Dk}{i(2\pi)^k}\frac{4k_\mu k_\nu}{\Big[k^2-m_1^2+i\varepsilon\Big]\Big[(k-p)^2-m_2^2+i\varepsilon\Big]}&=\frac{1}{\hat\epsilon}\left(m_1^2+m_2^2-\frac{p^2}{3}\right)\nonumber\\
       &\quad+4\bar{B}_{00}[p^2,m_1^2,m_2^2]\,,
\end{align}
Dimensional regularization is employed with $D=4-2\epsilon$ and $\mu$ is the renormalization scale. In the $\overline{\text{MS}}$ scheme, $1/\hat\epsilon=1/\epsilon -\gamma_E+\ln(4\pi)$. At $p^2=0$, the finite parts of these scalar integrals are,
\begin{align}
    \bar{A}_0[m^2]&=m^2\left(1-\ln\frac{m^2}{\mu^2}\right)\,,\\
    4\bar{B}_{00}[0,m_1^2,m_2^2]&=\bar{A}_0[m_1^2]+\bar{A}_0[m_2^2]+F[m_1^2,m_2^2]\,,\\
  4m_1^2\bar{B}_0[0,m_1^2,m_2^2]&=4\bar{A}_0[m_1^2]-2m_1^2-2m_2^2+4F[m_1^2,m_2^2]\,,
\end{align}
where the function $F[m_1^2,m_2^2]$ is defined as
\begin{equation}\label{eq:F-func}
F[m_1^2,m_2^2]=
\begin{cases}
0 & \text{if } \;m_1^2=m_2^2\,, \\
\frac{m_1^2+m_2^2}{2}-\frac{m_1^2m_2^2}{m_1^2-m_2^2}\ln\frac{m_1^2}{m_2^2} & \text{otherwise}.
\end{cases}
\end{equation}
\section{Most General Gauge-Invariant Lagrangian}\label{App:MostGenPotential}
In the SM extended by one real and one complex scalar triplet, the most general Higgs Lagrangian invariant under local $SU(2)_L\times U(1)_Y$, in the notation of~\cite{Kundu:2021pcg}, is given by
\begin{equation}\label{eq:genLag}
    \mathcal{L}_H=(D_\mu \phi)^\dagger(D^\mu \phi)+ (D_\mu \chi)^\dagger(D^\mu \chi)+\frac{1}{2}(D_\mu \xi)^\dagger(D^\mu \xi)-V(\phi,\chi,\xi)\,,
\end{equation}
where 
\begin{align}
V(\phi,\chi,\xi)&=-m_\phi^2\big(\phi^\dagger\phi\big)-m_\xi^2\big(\xi^\dagger\xi\big)-m_\chi^2\big(\chi^\dagger\chi\big)+\mu_1\big(\chi^\dagger t_a\chi\big)\xi_a+\mu_2\big(\phi^\dagger \tau_a\phi\big)\xi_a\nonumber\\
&\,\quad+\mu_3\Big[\big(\phi^T\epsilon\tau_a\phi\big)\tilde{\chi}_a+\text{h.c.}\Big]
+\lambda_\phi\big(\phi^\dagger\phi\big)^2+\lambda_\xi\big(\xi^\dagger\xi\big)^2+\lambda_\chi\big(\chi^\dagger\chi\big)^2\nonumber\\
&\,\quad+\tilde{\lambda}_\chi \big|\tilde{\chi}^\dagger\chi \big|^2+\lambda_{\phi\xi}\big(\phi^\dagger\phi\big)\big(\xi^\dagger\xi\big)
+\lambda_{\phi\chi}\big(\phi^\dagger\phi\big)\big(\chi^\dagger\chi\big)+\lambda_{\chi\xi}\big(\chi^\dagger\chi\big)\big(\xi^\dagger\xi\big)\nonumber\\
&\,\quad+\kappa_1\big|\xi^\dagger\chi \big|^2+\kappa_2\big(\phi^\dagger\tau_a\phi\big)\big(\chi^\dagger t_a\chi\big)+\kappa_3\Big[\big(\phi^T\epsilon\tau_a\phi\big)\big(\chi^\dagger t_a\xi\big)+\text{h.c.}\Big]\,.
\label{eq:v16}
\end{align} 
The scalar potential minimization conditions leading to the symmetry breaking $SU(2)_L\times U(1)_Y\to U(1)_{\text{EM}}$, are given by
\begin{align}\label{eq:tadpole-tree}
    0&=T_\phi=-m_\phi^2-\frac{1}{2}\mu_2v_\xi-\sqrt{2}\mu_3v_\chi+\lambda_\phi v_\phi^2+\left(\lambda_{\phi\chi}+\frac{\kappa_2}{2}\right)v_\chi^2+\lambda_{\phi\xi}v_\xi^2+\sqrt{2}\kappa_3v_\xi v_\chi\,,\nonumber\\
    0&=T_\xi=-2m_\xi^2-\mu_1\frac{v_\chi^2}{v_\xi}-\mu_2\frac{v_\phi^2}{4v_\xi}+4\lambda_\xi v_\xi^2+\lambda_{\phi\xi}v_\phi^2+2\lambda_{\chi\xi}v_\chi^2+\kappa_3\frac{v_\phi^2 v_\chi}{\sqrt{2}v_\xi}\,,\nonumber\\
    0&=T_\chi=-m_\chi^2-\mu_1v_\xi-\mu_3\frac{v_\phi^2}{2\sqrt{2}v_\chi}+2\lambda_\chi v_\chi^2+\frac{1}{2}\left(\lambda_{\phi\chi}+\frac{\kappa_2}{2}\right)v_\phi^2+\lambda_{\chi\xi}v_\xi^2+\kappa_3\frac{v_\phi^2 v_\xi}{2\sqrt{2}v_\chi}\,.
\end{align}
Imposing the $SU(2)_R$ symmetry on the scalar potential in Eq.~(\ref{eq:v16}) yields the following relations among the parameters:
\begin{align}\label{eq:GMconstraints}
m_\chi^2&=2m_\xi^2\,,&  \lambda_{\chi\xi}&=2\lambda_\chi-4\lambda_\xi\,,& \kappa_2&=4\lambda_{\phi\xi}-2\lambda_{\phi\chi}+\sqrt{2}\kappa_3\,,\nonumber\\
\mu_2&=\sqrt{2}\mu_3\,,& 
\kappa_2&=\sqrt{2}\kappa_3\,, & \kappa_1&=2\tilde{\lambda}_\chi=4\lambda_\xi-\lambda_{\chi\xi}\,.
\end{align}
Comparing Eqs.~(\ref{eq:v16}) and (\ref{GM_pot}) in the $SU(2)_R$ symmetric limit, we get
\begin{align}
    m_\phi^2&=-m_1^2\,, & m_\xi^2&=-\frac{1}{2}m_2^2\,, &  m_\chi^2&=-m_2^2\,, & \mu_1&=-6\tilde{\mu}_2\,,\nonumber\\
    \mu_2&=-\tilde{\mu}_1\,, & \mu_3&=-\frac{1}{\sqrt{2}}\tilde{\mu}_1\,, &  \lambda_\phi&=4\lambda_1\,, & \lambda_\xi&=\lambda_2+\lambda_3\,,\nonumber\\
    \lambda_\chi&=4\lambda_2+2\lambda_3\,, & \tilde{\lambda}_\chi&=2\lambda_3\,, &  \lambda_{\phi\xi}&=2\lambda_4\,, & \lambda_{\phi\chi}&=4\lambda_4\,,\nonumber\\
    \lambda_{\chi\xi}&=4\lambda_2\,, & \kappa_1&=4\lambda_3\,, &  \kappa_2&=2\lambda_5\,, & \kappa_3&=\sqrt{2}\lambda_5\,.
\end{align}
\section{Tadpole Renormalization}\label{App:tadpole}
\begin{figure}[htb!]
    \centering
    \includegraphics[width=0.3\textwidth]{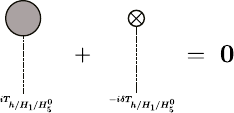}
    \caption{Tadpole renormalization conditions in the physical basis, where the tadpole counter-terms $\delta T_k$ ($k=h,H_1,H_5^0$) exactly cancel the sum of all one-loop tadpole contributions $T_k$ (shown in gray blob). }
    \label{fig:tadpole}
\end{figure}

In the mass eigenstate basis, the relevant terms of the most general gauge-invariant Lagrangian (see Eq.~(\ref{eq:genLag})) that generate tadpole diagrams, written in terms of bare parameters, are
\begin{align}\label{eq:BareLag}
    \mathcal{L}\supset&-\frac{v_\phi^0}{\sqrt{2}}\left[-(m_\phi^2)^0-\frac{1}{2}\mu_2^0v_\xi^0-\sqrt{2}\mu_3^0v_\chi^0+\Lambda_3^0(v^0)^2\right]\left(\frac{c_\alpha}{\sqrt{2}} h+\frac{s_\alpha}{\sqrt{2}} H_1\right)\\
    &-v_\xi^0\left[-(m_\xi^2)^0-\frac{\mu_1^0}{v_\xi^0}(v_\chi^0)^2-\frac{1}{4}\frac{\mu_2^0}{v_\xi^0}(v_\phi^0)^2+\Lambda_4^0(v^0)^2\right]\left(-\frac{s_\alpha}{\sqrt{3}} h+\frac{c_\alpha}{\sqrt{3}}H_1 -\sqrt{\frac{2}{3}}H_{5}^0\right)\nonumber\\
    &-v_\chi^0\left[-(m_\chi^2)^0-\mu_1^0v_\xi^0-\frac{1}{\sqrt{2}}\frac{\mu_3^0}{v_\chi^0}(v_\phi^0)^2+\Lambda_5^0(v^0)^2\right]\left(-\sqrt{\frac{2}{6}}s_\alpha h+\sqrt{\frac{2}{6}}c_\alpha H_1 +\sqrt{\frac{1}{6}}H_{5}^0\right)\,,\nonumber
\end{align}
where $(v^0)^2=(v_\phi^0)^2+8(v_\chi^0)^2$, and
\begin{align*}
    \Lambda_3^0&=\lambda_\phi^0\frac{(v_\phi^2)^0}{2(v^0)^2}+\left(\lambda_{\phi\chi}^0+\frac{\kappa_2^0}{2}\right)\frac{(v_\chi^2)^0}{(v^0)^2}+\lambda_{\phi\xi}^0\frac{(v_\xi^2)^0}{(v^0)^2}+\sqrt{2}\kappa_3^0 \frac{v_\chi^0v_\xi^0}{(v^0)^2}\,,\\
    \Lambda_4^0&=\lambda_\xi^0\frac{(v_\xi^2)^0}{(v^0)^2}+\lambda_{\phi\xi}^0\frac{(v_\phi^2)^0}{2(v^0)^2}+\lambda_{\chi\xi}^0\frac{(v_\chi^2)^0}{(v^0)^2}+\frac{\kappa_3^0}{\sqrt{2}}\frac{v_\chi^0}{v_\xi^0}\frac{(v_\phi^2)^0}{(v^0)^2}\,,\\
    \Lambda_5^0&=\lambda_\chi^0\frac{(v_\chi^2)^0}{(v^0)^2}+\frac{1}{2}\left(\lambda_{\phi\chi}^0+\frac{\kappa_2^0}{2}\right)\frac{(v_\phi^2)^0}{(v^0)^2}+\lambda_{\chi\xi}^0\frac{(v_\xi^2)^0}{(v^0)^2}+\frac{\kappa_3^0}{\sqrt{2}}\frac{v_\chi^0}{v_\xi^0}\frac{(v_\phi^2)^0}{(v^0)^2}\,.
\end{align*}
The right-hand side (RHS) of Eq.~(\ref{eq:BareLag}) vanishes at tree level due to the extremum conditions given in Eq.~(\ref{eq:tadpole-tree}). Beyond tree level, however, this cancellation does not hold in general. At one loop, the sum of all tadpole contributions is nonzero, $T_k \neq0$, for $k=h,H_1,H_5^0$. We therefore introduce the tadpole counter-terms $\delta T_h,\delta T_{H_1},$ and $\delta T_{H_5^0}$, which ensure that the renormalized tadpoles vanish order by order in perturbation theory. 
The corresponding renormalization conditions, illustrated in Figure~\ref{fig:tadpole}, are given by
\begin{equation}
    T_h-\delta T_h=0\,,\quad T_{H_1}-\delta T_{H_1}=0\,,\quad \text{and}\quad T_{H_5^0}-\delta T_{H_5^0}=0\,.
\end{equation}
The procedure then consists of omitting all tadpole contributions to the self-energies and mixing amplitudes. Note, however, that in theories without  custodial $SU(2)_V$ symmetry at tree-level, the relation $g_{\phi_iW^+W^-}\neq c_\theta^2 g_{\phi_iZZ}$ holds for $\phi_i=\tilde{h},\tilde{H_1},\tilde{H_5^0}$, corresponding to the CP-even scalars in the mass eigenstate basis. As a result, quadratic divergences appear in the tadpole contributions to $\Delta\rho^{\text{NS}}$ or equivalently the $T$ parameter given in Eq.~(\ref{eq:T-para}). The tadpole renormalization scheme described above is equally applicable in this case; consequently, $\Delta\rho^{\text{NS}}$ (or $T$ parameter) is free from power divergences. 
\section{Results for One-Loop Self-Energies and Mixings}\label{App:results}
The one-loop expressions of $H_5^+\text{--}G^+$ mixing (corresponding Feynman
diagrams are shown in Figure~\ref{fig:H5pGpMix}) are
\begin{align}
    16\pi^2\Sigma_{H_5^+G^+}^{(a)}&=\frac{e^2s_\beta}{2c_\theta^2s_\theta^2}\Big\{4c_\theta^2A_0[m_W^2]-4c_{2\theta}A_0[m_Z^2]+A_{\text{fin}}\Big\}\,,\\
    16\pi^2\Sigma_{H_5^+G^+}^{(b)}&=-\frac{e^2s_\beta}{4s_\theta^2}\Big\{2A_0[m_W^2]+2p^2B_0'[p^2,m_W^2,m_{H_5}^2]+B_0[p^2,m_{H_5}^2,m_W^2]\Big(m_{H_5}^2+3p^2\Big)\nonumber\\
&\quad\;-\frac{c_{2\theta}}{c_\theta^2}\Big(2A_0[m_Z^2]+2p^2B_0'[p^2,m_Z^2,m_{H_5}^2]+B_0[p^2,m_{H_5}^2,m_Z^2]\Big(m_{H_5}^2+3p^2\Big)\Big)\nonumber\\
&\quad\; +\frac{1}{2c_\theta^2}\Big(m_Z^2-\Big(2m_W^2-m_Z^2\Big)c_{2\theta}+\Big(3+c_{2\theta}\Big)p^2\Big)B_0[p^2,m_W^2,m_{Z}^2]
\nonumber\\
&\quad\; +\frac{1}{c_\theta^2}\Big(1+3c_{2\theta}\Big)p^2B_0'[p^2,m_W^2,m_{Z}^2]
\Big\}\,,\\
 16\pi^2\Sigma_{H_5^+G^+}^{(c)}&=-\frac{e^4v^2s_\beta}{2c_\theta^2s_\theta^2}\Big\{-2B_0[p^2,m_W^2,m_Z^2]+\text{finite terms}\Big\}\,,\\
 16\pi^2\Sigma_{H_5^+G^+}^{(d)}&=-\frac{e^4v^2s_\beta}{8c_\theta^2s_\theta^2}\Big\{B_0[p^2,m_W^2,m_Z^2]\Big\}\,,
\end{align}
where $B_0'\equiv dB_0/dp^2$, and $A_{\text{fin}}$ term contains no ultraviolet divergent contributions. The scalar functions $A_0$ and $B_0$ are defined in Appendix~\ref{App:PV-func}. For the $H_5^+\text{--}H_3^+$ mixing, the analytic expressions given below correspond to the Feynman diagrams shown in Figure~\ref{fig:fd-H5pH3p}. 
\begin{align}
16\pi^2\Sigma_{H_5^+H_3^+}^{(a)}&=\frac{e^2c_\beta}{2c_\theta^2s_\theta^2}\Big\{4c_\theta^2A_0[m_W^2]-4c_{2\theta}A_0[m_Z^2]+B_{\text{fin}}\Big\}\,,\\
 16\pi^2\Sigma_{H_5^+H_3^+}^{(b)}&=-\frac{e^2c_\beta}{4s_\theta^2}\Big\{2A_0[m_W^2]+2p^2\Big(B_0'[p^2,m_W^2,m_{H_3}^2]+B_0'[p^2,m_W^2,m_{H_5}^2]\Big)\nonumber\\
&\quad\;+B_0[p^2,m_{H_3}^2,m_W^2]\Big(m_{H_3}^2+3p^2\Big)+B_0[p^2,m_{H_5}^2,m_W^2]\Big(m_{H_5}^2+3p^2\Big)\nonumber\\
&\quad\;-\frac{c_{2\theta}}{c_\theta^2}\Big(2A_0[m_Z^2]+2p^2\Big(B_0'[p^2,m_Z^2,m_{H_3}^2]+B_0'[p^2,m_Z^2,m_{H_5}^2]\Big)\nonumber\\
&\quad\;+B_0[p^2,m_{H_3}^2,m_Z^2]\Big(m_{H_3}^2+3p^2\Big)+B_0[p^2,m_{H_5}^2,m_Z^2]\Big(m_{H_5}^2+3p^2\Big)\Big)\Big\}\,.
\end{align}
Here $B_{\text{fin}}$ term contains no ultraviolet divergent parts. 

For the gauge boson self-energies and mixings, the tadpole-free one-loop expressions given below correspond to the sum over all possible configurations of 
scalars, $S$ and $S'$ (see Figure~\ref{fig:fd-SVV}).
\begin{align}
16\pi^2\Sigma_{WW}^{(a)}&=-\frac{e^2}{4s_\theta^2}\Big\{\Big(\frac{11}{6}-\frac{5}{6}c_{2\alpha}\Big)A_0[m_h^2]+\Big(\frac{11}{6}+\frac{5}{6}c_{2\alpha}\Big)A_0[m_{H_1}^2]+\frac{40}{3}A_0[m_{H_5}^2]\nonumber\\
&\;\quad+2\Big(1+2s_\beta^2\Big)A_0[m_W^2]+\Big(1+s_\beta^2\Big)A_0[m_Z^2]+\Big(8-5s_\beta^2\Big)A_0[m_{H_3}^2]\Big\}\,,\\
16\pi^2\Sigma_{ZZ}^{(a)}&=-\frac{e^2}{4s_\theta^2c_\theta^2}\Big\{\Big(\frac{11}{6}-\frac{5}{6}c_{2\alpha}\Big)A_0[m_h^2]+\Big(\frac{11}{6}+\frac{5}{6}c_{2\alpha}\Big)A_0[m_{H_1}^2]+\Big(\frac{25}{3}+5c_{4\theta}\Big)A_0[m_{H_5}^2]\nonumber\\
&\;\quad+\Big(1+2s_\beta^2+c_{4\theta}\Big)A_0[m_W^2]+\Big(1+3s_\beta^2\Big)A_0[m_Z^2]+\Big(7-5s_\beta^2+c_{4\theta}\Big)A_0[m_{H_3}^2]\Big\},\\
16\pi^2\Sigma_{\gamma Z}^{(a)}&=-\frac{e^2(c_\theta^2-s_\theta^2)}{c_\theta s_\theta}\Big\{5A_0[m_{H_5}^2] +A_0[m_{W}^2]+A_0[m_{H_3}^2]\Big\}\,,\\
16\pi^2\Sigma_{\gamma\gamma}^{(a)}&=-2e^2\Big\{5A_0[m_{H_5}^2] +A_0[m_{W}^2]+A_0[m_{H_3}^2]\Big\}\,,
\end{align}
where $A_{0}$ function is defined in Appendix~\ref{App:PV-func}.
\begin{align}
16\pi^2\Sigma_{WW}^{(b)}&=\frac{e^2}{s_\theta^2}\Big\{ \frac{10}{3}c_\beta^2 B_{00}[p^2,m_{H_3}^2,m_{H_5}^2]+\frac{7}{3}s_\beta^2 B_{00}[p^2,m_{W}^2,m_{H_5}^2]+5B_{00}[p^2,m_{H_5}^2,m_{H_5}^2]\\\nonumber
&\quad+B_{00}[p^2,m_{H_3}^2,m_{H_3}^2]+B_{00}[p^2,m_{W}^2,m_{Z}^2]+s_\beta^2B_{00}[p^2,m_{Z}^2,m_{H_5}^2]\\\nonumber
&\quad +\Big(c_\alpha s_\beta+\sqrt{\frac{8}{3}}s_\alpha c_\beta\Big)^2B_{00}[p^2,m_{H_3}^2,m_{h}^2]+\Big(s_\alpha s_\beta-\sqrt{\frac{8}{3}}c_\alpha c_\beta\Big)^2B_{00}[p^2,m_{H_3}^2,m_{H_1}^2]\\\nonumber
&\quad +\Big(c_\alpha c_\beta-\sqrt{\frac{8}{3}}s_\alpha s_\beta\Big)^2B_{00}[p^2,m_{W}^2,m_{h}^2]+\Big(s_\alpha c_\beta+\sqrt{\frac{8}{3}}c_\alpha s_\beta\Big)^2B_{00}[p^2,m_{W}^2,m_{H_1}^2]\Big\}\,,\\
16\pi^2\Sigma_{ZZ}^{(b)}&=\frac{e^2}{s_\theta^2c_\theta^2}\Big\{\frac{4}{3}s_\beta^2 B_{00}[p^2,m_{Z}^2,m_{H_5}^2]+2s_\beta^2B_{00}[p^2,m_{W}^2,m_{H_5}^2]+c_{2\theta}^2B_{00}[p^2,m_{W}^2,m_{W}^2]\\\nonumber
&\quad+\frac{10}{3}c_\beta^2 B_{00}[p^2,m_{H_3}^2,m_{H_5}^2]+c_{2\theta}^2B_{00}[p^2,m_{H_3}^2,m_{H_3}^2]+5c_{2\theta}^2B_{00}[p^2,m_{H_5}^2,m_{H_5}^2]\\\nonumber
&\quad +\Big(c_\alpha s_\beta+\sqrt{\frac{8}{3}}s_\alpha c_\beta\Big)^2B_{00}[p^2,m_{H_3}^2,m_{h}^2]+\Big(s_\alpha s_\beta-\sqrt{\frac{8}{3}}c_\alpha c_\beta\Big)^2B_{00}[p^2,m_{H_3}^2,m_{H_1}^2]\\\nonumber
&\quad + \Big(c_\alpha c_\beta-\sqrt{\frac{8}{3}}s_\alpha s_\beta\Big)^2B_{00}[p^2,m_{Z}^2,m_{h}^2]+\Big(s_\alpha c_\beta+\sqrt{\frac{8}{3}}c_\alpha s_\beta\Big)^2B_{00}[p^2,m_{Z}^2,m_{H_1}^2]\Big\}\,,\\
16\pi^2\Sigma_{\gamma Z}^{(b)}&=\frac{2e^2(c_\theta^2-s_\theta^2)}{c_\theta s_\theta}\Big\{5B_{00}[p^2,m_{H_5}^2,m_{H_5}^2]
+B_{00}[p^2,m_{H_3}^2,m_{H_3}^2]+B_{00}[p^2,m_{W}^2,m_{W}^2]\Big\}\,,\\
16\pi^2\Sigma_{\gamma\gamma}^{(b)}&=4e^2\Big\{5B_{00}[p^2,m_{H_5}^2,m_{H_5}^2]
+B_{00}[p^2,m_{H_3}^2,m_{H_3}^2]+B_{00}[p^2,m_{W}^2,m_{W}^2]\Big\}\,,
\end{align}
where $B_{00}$ function is defined in Appendix~\ref{App:PV-func}.
\begin{align}
16\pi^2\Sigma_{WW}^{(c)}&=-\frac{e^2m_W^2}{s_\theta^2}\Big\{\frac{7}{3}s_\beta^2 B_{0}[p^2,m_{W}^2,m_{H_5}^2]+ \Big(c_\alpha c_\beta-\sqrt{\frac{8}{3}}s_\alpha s_\beta\Big)^2B_{0}[p^2,m_{W}^2,m_{h}^2]\nonumber\\
&\;\quad +\frac{s_\beta^2}{c_\theta^2} B_{0}[p^2,m_{Z}^2,m_{H_5}^2]+\Big(s_\alpha c_\beta+\sqrt{\frac{8}{3}}c_\alpha s_\beta\Big)^2B_{0}[p^2,m_{W}^2,m_{H_1}^2]\nonumber\\
&\;\quad+ \frac{s_\theta^4}{ c_\theta^2}B_{0}[p^2,m_{W}^2,m_{Z}^2]+s_\theta^2 B_{0}[p^2,0,m_{W}^2] \Big\}\,,
\end{align}
\begin{align}
16\pi^2\Sigma_{ZZ}^{(c)}&=-\frac{e^2m_Z^2}{s_\theta^2c_\theta^2}\Big\{\frac{4}{3}s_\beta^2 B_{0}[p^2,m_{Z}^2,m_{H_5}^2]+ \Big(c_\alpha c_\beta-\sqrt{\frac{8}{3}}s_\alpha s_\beta\Big)^2B_{0}[p^2,m_{Z}^2,m_{h}^2]\nonumber\\
&\;\quad +2s_\beta^2c_\theta^2 B_{0}[p^2,m_{W}^2,m_{H_5}^2]+\Big(s_\alpha c_\beta+\sqrt{\frac{8}{3}}c_\alpha s_\beta\Big)^2B_{0}[p^2,m_{Z}^2,m_{H_1}^2]\nonumber\\
&\;\quad+2s_\theta^4 c_\theta^2B_{0}[p^2,m_{W}^2,m_{W}^2]\Big\}\,,\\
16\pi^2\Sigma_{\gamma Z}^{(c)}&=-\frac{2e^2s_\theta }{c_\theta} m_W^2B_{0}[p^2,m_{W}^2,m_{W}^2]\,,\\
16\pi^2\Sigma_{\gamma\gamma}^{(c)}&=-2e^2m_W^2 B_{0}[p^2,m_{W}^2,m_{W}^2]\,,
\end{align}
where $B_{0}$ function is defined in Appendix~\ref{App:PV-func}.
\bibliography{refs}
\end{document}